# Mass-sensing BioCD Protein Array towards Clinical Application: Prostate Specific Antigen Detection in Patient Sera

Xuefeng Wang[1], Ming Zhao[1] David D. Nolte[1] and Timothy L Ratliff[2]

1 Department of Physics, 525 Northwestern Avenue, West Lafayette, IN 47907-2036

Corresponding author: nolte@physics.purdue.edu

2 Purdue Cancer Center, Hansen Life Sciences Research Building, West Lafayette, IN 47907-2064

**Abstract:** Mass-sensing biosensor arrays for protein detection require no fluorophores or enzyme labels. However, few mass biosensor protein arrays have demonstrated successful application in high background samples, such as serum. In this paper, we test the BioCD as a mass biosensor based on optical interferometry of antibodies covalently attached through Schiff-base reduction. We use the BioCD to detect prostate specific antigen (PSA, a biomarker of prostate cancer) in patient sera in a 96-well anti-PSA microarray. We have attained a 4 ng/ml detection limit in full serum and have measured PSA concentrations in three patient sera.

## Introduction

The development of mass-sensing biosensors (without the need for fluorophore labeling, radiolabels or other chemicals) for protein detection is advancing rapidly. Surface plasmon resonance (SPR) [1, 2], imaging ellipsometry [3, 4], photonic gratings [5], BioCD [6-9], nanowires [10], and micro-cantilevers [11, 12] have been proposed and demonstrated in succession. These biosensors sense the mass of biomaterial and have the potential for high-throughput and quantitative protein detection. Methods such as Western blot or enzyme-linked immunosorbent assay (ELISA) detect specific protein via fluorescence or other chemically-induced signals. These signals are assumed to be proportional to the amount of target protein. However, the intensity of fluorescence and the chemically-induced signals may be adversely affected by other parameters such as the local electrical field intensity, pH, temperature, or may be subject to bleaching and quenching. On the other hand, mass biosensors measure protein quantity by directly converting the protein mass into physical signals which are more independent of the local micro-environment. For example, SPR, ellipsometry and the BioCD convert protein mass into optical quantities (surface plasmon, polarization and light intensity, respectively). Nanowires and micro-cantilevers convert protein mass into electrical and mechanical quantities, respectively. Among these mass biosensors, the BioCD converts protein mass on a $SiO_2$/Si wafer into a reflectance variation via local interferometry which optimizes the protein signal under a quadrature condition [6]. The large area of the wafer can accommodate thousands of assays simultaneously. $SiO_2$/Si wafers can be obtained economically, and the $SiO_2$ modification has been extensively studied for protein immobilization. In addition, the BioCD scanning system adopts a spinning-disc approach which reduces the detection noise floor and improves protein detection limits [13]. The detection resolution for average protein layer thickness can be as small as 2 pm for a single 150 micron-diameter protein spot.

Protein array mass biosensors face obstacles in clinical applications because the background protein level in serum is much higher than the target protein concentration and induces strong non-specific binding. Non-specific binding can mask the target signal because label-free biosensors detect all protein present rather than only the specific target protein. In this report, we adopted an enhanced Schiff-base method to produce butyraldehyde-modified silicon wafers for an antibody microarray that significantly suppresses the non-specific binding. The high sensitivity of the BioCD combined with stable surface chemistry allows us to detect the target protein at ng/ml levels in a high protein background, such as serum (total protein level >40 mg/ml [14]).

**Experimental Methods**

**Principle of the BioCD.** The BioCD has the format of a disc with dielectric coatings. The coatings create a surface reflection coefficient $r$ and reflectance $R=|r|^2$. After a protein layer is immobilized on the coating surface, the reflectance is changed to $R'$. We have shown previously [15] that the reflectance increment caused by protein, $\Delta R = R'-R$, has an explicit relation to the original $r$, expressed by

$$\Delta R = \text{Im}\left[(1+r)^2 \bar{r}\right](n_p^2 - 1)\frac{2\pi d}{\lambda \cos\theta_0} \tag{1}$$

where $n_p$ is the refractive index of the protein layer, $d$ is the thickness of the protein layer, $\lambda$ is the wavelength of the probe light, and $\theta_0$ is the incident angle. A laser scans the disc, measuring the height (or mass area density) of printed protein. For a defined protein layer and probe wavelength, where $n_p$, $d$, $\lambda$ and $\theta_0$ are constants, $\Delta R$ is maximized when $r = \pm 0.577i$. A silicon wafer with 120 nm of thermal SiO$_2$ has a calculated $r = -0.049 + 0.489i$ at a wavelength of 488 nm with s-polarization and 30° incidence, which is near the optimal value for $r$. This working condition has only one dielectric layer (SiO$_2$). On this 120 nm SiO$_2$ wafer, a 1 nm protein layer induces 0.0056 absolute reflectance change, or 2.3% relative intensity increment. We use a spinning disc scanning system to acquire the reflectance map of the entire disc, which is converted into a map of protein areal density or protein layer height.

**Scanning system.** The experimental layout of the BioCD scanning system is shown in Fig. 1. The probe light is incident at 30º on the disc with s-polarization, although unpolarized normally-incident light works as well because interferometry does not require polarization analysis. The light source is an INNOVA 300 Argon laser (Coherent, Inc.) using a wavelength of 488 nm. The reflectance signal of the BioCD is detected by a silicon-based detector. The BioCD is mounted on a motor (Lincoln Laser, Inc.) on a linear translation stage (MM2000, Newport Corp). 2D scanning is realized by spinning the motor and moving the stage. A computer controls the entire system, collects the reflectance data and reconstructs the reflectance map of the entire BioCD with a resolution of 20 μm. One scan of a 100 mm diameter BioCD is completed in one hour. The reflectance map is converted to a protein thickness profile using the conversion ratio $d\ (nm) = \Delta R / R / 0.023$, where $R$ is the background reflectance, and $\Delta R$ is the reflectance change caused by the protein layer.

**Surface Chemistry.** The 120 nm thermal oxide on silicon was aldehyde-functionalized to bind protein covalently. Butyraldehyde was grafted on the $SiO_2$ surface through silanization where the silanol group (Si-OH) on the aldehydic silane reacts with a hydroxyl group (-OH) on hydrated silica surface and forms a disiloxane bond(Si-O-Si). The antibodies are printed on the functionalized silica surface by a protein printer. The carbonyl group of butyraldehyde reacts with the amino groups of the antibody to form a Schiff base [16], a carbon-nitrogen double bond which immobilizes the antibody molecules through primary amines. After antibody printing, the surface was reduced by a $NaBH_4$ solution which stabilizes antibody attachment by converting carbon-nitrogen double bond (C=N) to carbon-nitrogen single bond (C-N). The $NaBH_4$ also helps to block the substrate by reducing remnant carbonyl groups to hydroxyl groups. The BioCD is further blocked by casein solution (Phosphate buffered saline (PBS) + 1% casein solution in 20:1 volume ratio, PH 8.0), washed by 50 mM citric acid solution (PH 6.0), PBST (PBS + 0.05% Tween 20) and deionized water and spun dry at 1000 rpm. Bovine serum albumin (BSA) is widely used for surface blocking in many antibody microarrays, but it is not recommended for the mass-sensing BioCD because BSA is a large protein molecule which contributes more mass noise for interferometric biosensors. Here, we use casein for its small molecular weight.

**Prostate Specific Antigen (PSA).** We applied the BioCD to detect PSA levels in prostate cancer patient sera. PSA is a member of the kallikrein family exclusively produced by the prostate gland. It has a concentration of 0.5 mg/mL to 2.0 mg/mL in normal seminal fluid [17, 18] and 0 to 4 ng/ml [19] in normal serum. Prostate cancer and other prostate pathological changes may cause PSA to leak into the peripheral circulation and induce an abnormal elevation of PSA in serum. Levels higher than 4 ng/ml may suggest the presence of prostate cancer. Therefore PSA can work as a biomarker for prostate cancer. Measurement of PSA in serum is one of the standard diagnostics for prostate cancer and is usually performed by ELISA. There has been no previous study showing successful PSA measurement in patient sera based on mass-sensing biosensors in a microarray format.

**96-well BioCD.** We print antibody and reference protein in the format of paired target and reference spots on the BioCD and perform a prescan to record the initial height of all protein spots. During analyte solution incubation, antibodies capture target antigen due to immuno-binding, and the height of the antibody spot increases. A postscan is performed to find the protein spot heights after the assay. The specific height increment is defined as the antibody increment minus the reference-spot increment. If nonspecific binding is similar to both spots, then this procedure subtracts this background contribution. The difference is then related to the antigen concentration in the sample. Together—printing antibody, a prescan, incubation of analyte solution followed by a second specific antibody, and a postscan—complete one assay.

To perform multiple assays simultaneously on a single BioCD, the surface of a 100 mm diameter disc is separated into 96 wells with ultra-hydrophobic surface pads shown in Fig. 2 a), b) and c). The contact angle of water on the pads is as large as $90^o$ to confine the incubation solution within each well, and 96 assays can be performed in parallel on a single disc. The 96-well BioCD enables multiple sample tests or multiplexed protein detection, respectively, by incubating with different samples or by printing multiplexed antibodies in wells. Fig. 2 shows the prescan of a 96-well anti-PSA BioCD used in the following experiment. The protein mass profile in thickness is derived from the reflectance map and recorded for each well shown in Fig. 2d). An assay example is shown in Fig. 3 in which a negative and a positive response were observed, respectively, for 0 and 1 μg/ml analyte incubation.

**PSA Detection in Patient Sera.** We applied the 96-well anti-PSA BioCD to PSA detection in patient sera. A Scienion (BioDot Corp.) piezoelectric printer was used to print 32 antibody spots (affinity purified polyclonal anti-PSA produced in goat, G-126-C, Biospacific Co.) and 32 reference spots (anti-rabbit IgG produced in goat, R2004, Sigma Inc.) in each of the 96 wells on a butrylaldehyde-activated BioCD. The diameter of each spot was approximately 150 μm with a spot-to-spot separation of approximately 300 μm. Each antibody spot is locally paired with a reference spot in a format as shown in Fig. 2 d). A prescan was performed to record the original protein profile of the entire BioCD after printing.

Test samples included: 1) female serum (PSA level is null in female serum so it is a negative control), 2) PSA spiked into female serum, and 3) prostate-cancer patient sera. The three patient sera were acquired from the Purdue Cancer Center. The PSA concentrations had been measured with ELISA to be 34, 117 and 2138 ng/ml respectively for samples 1, 2 and 3. Female serum was purchased from Innovative Research, Inc. Affinity purified PSA reagent was purchased from Fitzgerad Co.

We created a series of PSA-spiked female sera with PSA concentrations 0, 4, 12.6, 40, 126, 400, 1260, 4000, and 12640 ng/ml. The PSA-spiked serum was used to generate a standard PSA response curve. For all three patient sera, we performed a series dilution in half-logarithmic sequential ratios 1:1, 1:3.16, 1:10, 1:31.6, 1:100 diluted by female serum (the ratio 1:m is defined as 1 patient serum + m-1 female serum). Dilution with female serum instead of buffer solution maintains an equal protein background for all test samples.

Prior to incubation, all samples were diluted by PBST solution in a ratio 1:4. This step is crucial because Tween-20 in the buffer suppresses non-specific binding. All samples shared the same protein background level (i.e., 1/4 of full human serum) at about 10 mg/ml. Patient sera dilution ratios then became 1:4, 1:12.6, 1:40, 1:126 and 1:400. PSA concentrations in spiked female serum became 0, 1, 3.16, 10, 31.6, 100, 316, 1000, 3160 ng/ml. We incubated the 96-well anti-PSA BioCD with these prepared samples. Each sample was distributed across 4 randomly allotted wells to suppress local bias on the disc. Overall, 9×4 wells were used for PSA spiked serum to acquire the standard response curve and 3×5×4 wells were used for diluted patient sera to acquire the curves for sequential dilution. The incubation time was 1 hour, after which the BioCD was washed for 5 min with PBST and 5 min with DI water (on an orbital shaker) and dried with pure nitrogen. After this procedure, PSA molecules in the samples had been captured by the anti-PSA spots on the BioCD.

Sandwich assays were performed to amplify the assay response through the added mass of the second antibody. All 96 wells were incubated with 10 μg/ml anti-PSA (G-126-C, Biospacific Co.) in PBST for 30 min (static incubation), followed by 5 min PBST and 5 min DI water wash (on orbital shaker), and dried with pure nitrogen. The anti-PSA antibodies bind with the captured PSA antigen and amplify the antibody spot height increment in the assays. A postscan was performed to record all protein spot heights after the sandwich incubation.

## Results and Discussions

The immunoassay responses in all wells were computed from the prescan and postscan data of the 96-well BioCD reflectance map (with a resolution of 20 μm). An image analysis program automatically located all wells on the map, extracted protein spot profiles for each well and calculated the pre- and post-height of all 6144 spots. The assay response was then calculated for each pair of antibody and reference spots by the algorithm $(postH-preH)_{antibody} - (postH-preH)_{reference}$. The analysis can be completed in less than 10 minutes on a desktop computer.

**Evaluation of the Detection Limit for PSA.** We evaluated the limit of detection (LOD) for PSA by comparing the assay responses for 0, and 1 ng/ml PSA spiked sera presented in Fig. 4a (LOD evaluation). The error bars represent the standard error of the assay variation for each test sample using 128 spots as the spot population number (4 replicate wells). The assay response is 0.007 nm for 0 ng/ml PSA spiked female serum as a negative control sample and 0.039 nm for 1 ng/ml PSA spiked female serum. The standard errors of the tests are respectively 0.0094 nm and 0.0078 nm. Using a 3σ criterion for a positive response, 0.039-0.007 = 0.032 (nm) is larger than three times the standard error 0.0094 nm, making the assay response positive at 1 ng/ml in PSA-spiked female serum (1:4 dilution of full serum) with a background of about 10 mg/ml. The detectable PSA level is 0.1 ppm of the background concentration. Therefore, the 1 ng/ml detection limit in 1:4 serum is equivalent to 4 ng/ml in full serum, which is the clinical threshold for prostate cancer diagnosis.

**PSA Concentration recovery in patient sera.** A standard PSA response curve was generated from the assays of PSA-spiked female sera. The dilution curves of patient samples were acquired from sequentially-diluted samples. We measured the PSA levels in the patient sera by matching the effective concentrations of the dilution series

Both the standard response and the dilution curves are fit by the Langmuir equation [20]

$$H = H_{Satur} \frac{[PSA]}{[PSA]+K_D} \quad \text{and} \quad H = H_{Satur} \frac{D}{D+D_C}$$

where D is the dilution ratio, $K_D$ is the equilibrium constant, and $D_C$ is the central dilution ratio at which the assay response is the half of the saturated response. $K_D$ and $D_C$ are calculated by fitting the standard response curve and the dilution curves of the patient samples. Because $D_C$ is the dilution ratio at which the diluted sample has the same assay response as a sample with an analyte concentration equal to $K_D$, the analyte concentration in the test sample is recovered as the ratio $K_D / D_C$.

The standard response curve and patient sera dilution curves are shown in Fig. 4 a) and b). $K_D$ was found to be 15.8 ng/ml, and the $D_C$ are 0.51, 0.32 and 0.0033, respectively, for patient serum samples 1, 2 and 3. The recovered PSA concentrations are 15.8/0.51 = 31 (ng/ml) for sample 1, 15.8 /0.32 = 49 (ng/ml) for sample 2 and 15.8/0.0033 = 4788 (ng/ml) for sample 3. PSA levels in the three samples were measured previously as 34, 117 and 2138 ng/ml by ELISA.

## Conclusions

We have demonstrated a mass-sensing biosensor capable of measuring a low-level cancer biomarker in patient sera. Mass-sensing biosensor protein arrays have been challenged by strong non-specific binding caused by the high protein background in serum. Currently, few mass-sensing biosensor protein arrays successfully detect target protein in serum or other body fluids and therefore are excluded from clinical application. To suppress the non-specific binding, we used a reference-spot subtraction analysis plus Schiff-base reduction and casein blocking. For incubation, we added 0.05% Tween 20 into analyte solution to further suppress non-specific binding. Moreover, the locally-paired reference spot for each antibody spot helped to monitor and neutralize the local non-specific binding. The large population of antibody spots further reduced the detection error. The combination of these factors has allowed us to successfully perform PSA detection in prostate cancer patient sera. We found a 1 ng/ml PSA detection limit in PSA-spiked serum with 10 mg/ml background. This detection limit is equivalent to 4 ng/ml PSA in full serum ,which is the threshold for normal-to-abnormal PSA levels. We also recovered the PSA concentrations for three patient serum samples at 31, 49 and 4788 ng/ml compared with 34, 117 and 2138 ng/ml measured by traditional enzyme-labelled ELISA. The results suggest that the combination of the BioCD interferometric laser scanning with the enhanced Schiff base chemistry has the potential for clinical applications.

## Acknowledgement

This work was sponsored under grants from Quadraspec, Inc. and from the Indiana Economic Development Corporation through the Purdue Research Foundation.

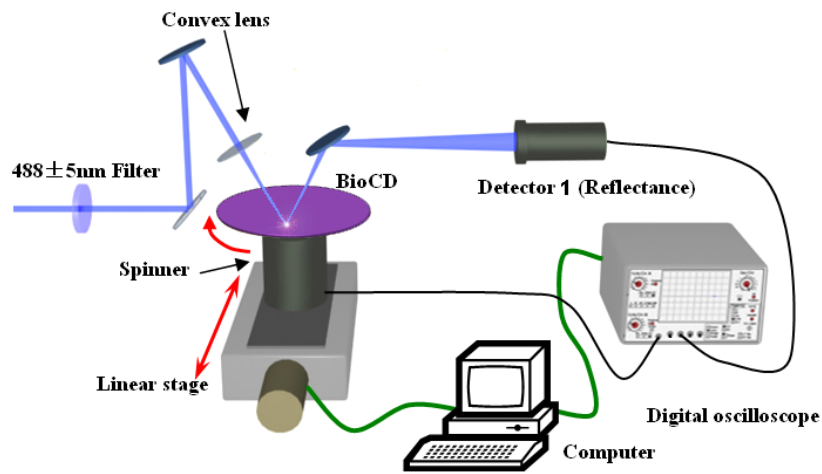

Fig. 1. BioCD Scanning system. This system scans the BioCD and acquires a 2D reflectance map with 20 μm resolution. Probe light with 488 nm wavelength is obliquely incident to achieve an optimal interferometric condition for protein sensing.

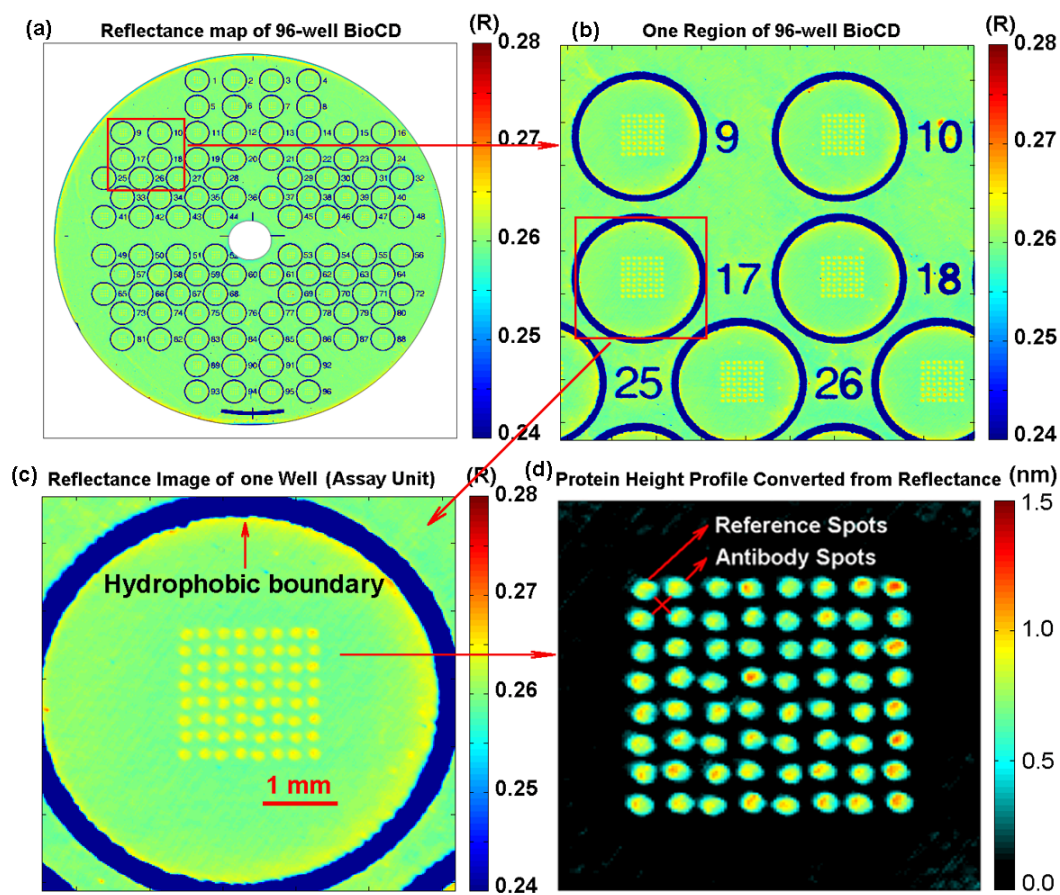

Fig.2. Imaging of protein profiles on a 96-well BioCD.  A) and B) 96 wells were isolated and numbered with ultra-hydrophobic ink on a thermal-oxide silicon wafer with 120 nm $SiO_2$.  C) Antibody and reference spots were printed in wells in an 8x8 spot pattern consisting of 2x2 unit cells of two target and two reference spots. Incubation solution is confined within each single well by the hydrophobic boundary allowing multiple sample tests performed on a single BioCD.  D) Protein spot mass profile (in terms of protein layer height in units of nm) of each well is derived from the reflectance map.

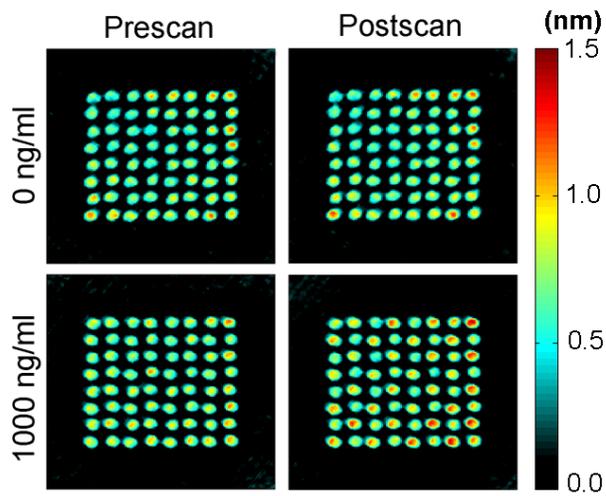

Fig.3. Demonstration of two well-based sandwich assays at 0 ng/ml and 1000 ng/ml. A prescan was performed before the assays to record the original height of the protein spots. After the sandwich assays, a postscan recorded the new protein heights. For the well incubated with 0 ng/ml PSA spiked serum, the average specific increment is 0.007 nm, and for the well incubated by 1000 ng/ml PSA spiked serum, the increment is 0.18 nm.

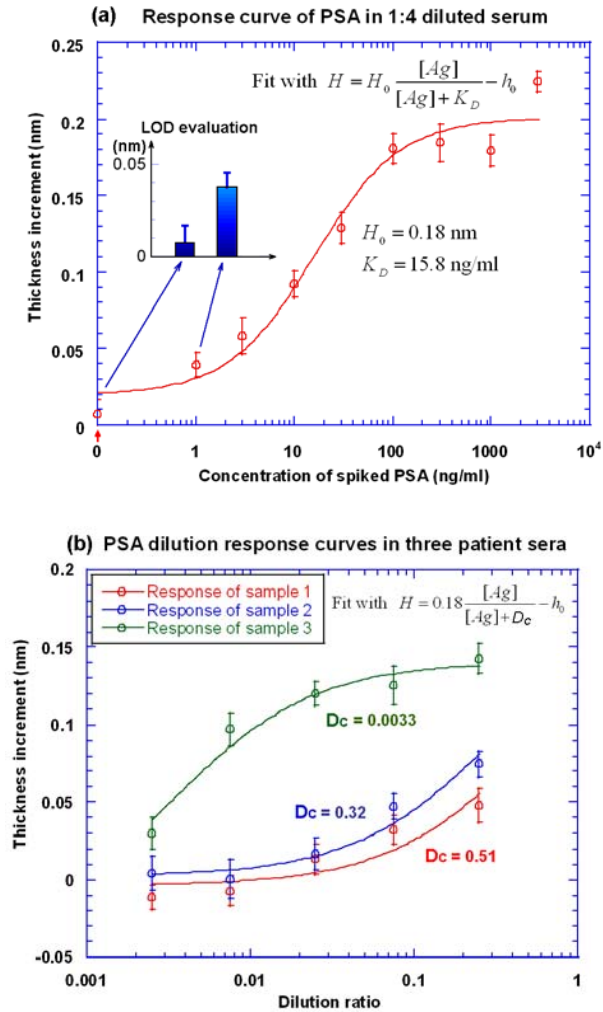

Fig.4. PSA concentration recovery for three patient samples.  We compared and matched the dilution curves shown in Fig. b) with the standard response curve in Fig. a) to find the concentration of PSA in the patient sera.  The standard curve was acquired from the assay results based on a series of PSA-spiked female sera and fit by a Langmuir equation.  The equilibrium constant $K_D$ was 15.8 ng/ml for PSA in serum.  The central dilution ratios $D_C$ were 0.51, 0.32, and 0.0033 for the three patient samples.  The PSA concentrations are calculated as [PSA]=$K_D/D_C$ and are 31, 49, 4788 ng/ml respectively for sample 1, 2 and 3.  The PSA levels measured by ELISA in these three samples were 34, 117 and 2138 ng/ml.